\documentclass[sigconf,nonacm]{acmart}

\renewcommand\footnotetextcopyrightpermission[1]{}
\usepackage{multirow}

\begin{document}

\title{UniCASE: A Unified 16-bit Floating-Point Format with Criticality-Aware Selective ECC for Efficient DNN Protection}




\author{Amna Hassan}
\affiliation{%
  \institution{University of Amsterdam}
  \city{Amsterdam}
  \country{The Netherlands}
}
\email{a.hassan@uva.nl}

\author{Semeen Rehman}
\affiliation{%
  \institution{University of Amsterdam}
  \city{Amsterdam}
  \country{The Netherlands}
}
\email{s.rehman@uva.nl}

\begin{abstract}
Soft errors are an increasing reliability concern for Deep Neural Network execution because they can corrupt parameters, leading to accuracy degradation. While conventional ECC offers strong fault protection, it incurs additional parity storage and computational overhead. Embedded-parity formats reduce storage cost by reusing the least-significant bits, but they do not optimize protection while reducing computational overhead. We propose UniCASE, a unified 16-bit floating-point (FP) format that jointly optimizes data representation and error protection for reliable DNN execution. It identifies stable blocks across FP64, FP32, FP16, and BFloat16 that can be mapped into a unified representation. Based on bit-level criticality analysis, UniCASE uses selective ECC that assigns distinct levels of protection to different data bits according to their resilience against soft errors. Experimental results show that UniCASE reduces encoder/decoder cost by up to \(30\%\), preserves model accuracy within \(1\%\) of the FP32 baseline, and provides significantly stronger soft error resilience than existing embedded-parity methods.
\end{abstract}




\maketitle

\begingroup
\renewcommand\thefootnote{}
\footnotetext{\textit{Author-prepared preprint.} This paper has been accepted for publication at ASP-DAC 2027. The final version of record will appear in the ACM Digital Library. DOI: 10.1145/3840404.3848334.}
\endgroup

\section{Introduction}
Soft errors pose a growing threat to reliable Deep Neural Network (DNN) execution, as transient bit flips in memory can corrupt weights, activations, or intermediate data and lead to incorrect predictions or misclassification \cite{li2017understanding}. This threat is amplified as DNNs have scaled by several orders of magnitude over the last decade \cite{villalobos2022machine}. The growing number of parameters and activations stored in DRAM, SRAM buffers, caches, and register files increases the number of vulnerable memory bits that are exposed to soft errors during execution \cite{baumann2005soft}. A typical 45-nm 6T-SRAM cell can experience approximately 1095 errors per million cells per billion hours \cite{vijayan2017online}, while a production-scale study observed 75.1 million DRAM errors over eight months across 250K servers \cite{cheng2022depth}. These observations show that SRAM and DRAM faults are practical reliability concerns for memory-intensive DNN execution.
To protect memory units against soft errors, error correcting code (ECC) has been used as a standard protection mechanism. However, it comes with non-trivial overhead, as ECC requires additional parity-bit storage as well as encoding and decoding logic. For example, conventional ECC DRAM increases raw module capacity by 12.5\% to store reliability metadata \cite{luo2017using}, while recent DRAM ECC schemes can require several to tens of decoding cycles depending on correction strength \cite{bae2023twin}. As DNN memory footprints continue to grow, protecting all weights and activations with ECC becomes increasingly expensive. \par
To reduce the storage and computational overhead of conventional ECC, prior studies \cite{wang2021ftapprox,mishra2023vadf,gracia2024allocating,mishra2025sera} have explored embedding parity bits directly within the data format by exploiting the error tolerance and insignificant bits present in deep-learning data formats. These approaches eliminate the need for separate ECC storage by reusing low-impact bits for protection metadata. However, such embedded-parity methods provide limited correction capability and may not offer the same protection coverage as conventional ECC. Moreover, they overlook that DNN resilience does not require storing or protecting all data bits with full precision, and not all data bits contribute equally to model correctness and therefore do not require the same level of protection. As a result, existing methods do not fully exploit ECC design flexibility to reduce parity-bit count and computation while preserving strong protection for the most critical DNN data. \par
To address these limitations, we propose UniCASE, a compressed 16-bit data format unified across FP64, FP32, FP16, and BFloat16 that preserves only the most computationally critical information required for reliable DNN execution. UniCASE classifies data bits into different criticality levels through fault-injection analysis and assigns protection strength according to their impact on final model accuracy.
Building on this representation, we introduce a criticality-aware ECC scheme for DNNs that achieves classification accuracy comparable to conventional SEC-DAEC-TAEC ECC while using fewer parity bits and reducing encoder/decoder XOR cost by up to 30\%. By exploiting the inherent bit-level error tolerance of DNN parameters, the proposed compression incurs zero to minimal accuracy loss while substantially lowering computational overhead. Moreover, the format is designed to be generalizable across model classes, making it applicable not only to conventional DNNs but also to larger and more complex models such as LLMs. Our primary contributions are summarized as follows:
\vspace{-5mm}
\begin{itemize}
\item We propose a compressed 16-bit format that retains only computation-critical DNN parameter bits to integrate parity bits within the same 16-bit word, while keeping accuracy loss within 1\% of the FP32 baseline (see Section \ref{analysis} and \ref{acc_loss}).
\item We present a cross-format bit profiling analysis across FP64, FP32, FP16, and BFloat16 to identify stable exponent and mantissa bits across these data formats that can be mapped into a common representation (see Section \ref{analysis}).
\item We introduce a criticality-aware multi-level ECC scheme that selectively assigns stronger protection to accuracy-sensitive bits and avoids unnecessary protection for low-impact bits (see Section \ref{CriticalityAware}).
\item We develop a five-parity-bit syndrome assignment method that provides selective SEC/DAEC/TAEC protection while reducing  ECC encoding and decoding cost by up to 30\% (see Section \ref{CriticalityAware} and \ref{comparison}).
\end{itemize}
\section{Related Work}
Conventional ECCs are commonly used to protect neural-network parameters stored in AI accelerator memories against soft errors \cite{raji2025ecc,traiola2023hardnning,feinberg2018making}. \cite{raji2025ecc} use stored ECC to detect parameter bit flips and correct or mask faulty weights during inference. ~\cite{traiola2023hardnning} use fault analysis to selectively protect critical DNN parameters with ECC. Feinberg et al.~\cite{feinberg2018making} apply error correction to improve the reliability of memristive neural-network accelerators. However, these ECC-based methods uniformly protect all data bits and require dedicated parity storage together with increasing encoding and decoding logic complexity, thus resulting in significant storage and computational overhead.

To reduce the storage overhead of ECC, recent studies \cite{wang2021ftapprox,mishra2023vadf,gracia2024allocating,mishra2025sera} have explored embedding parity information directly into the floating-point format. FTApprox \cite{wang2021ftapprox} improves soft-error resilience through an approximate data format, but it only provides single error detection. VADF~\cite{mishra2023vadf} improves robustness under single-bit-flip scenarios, but it does not provide strong correction guarantees for multi-bit faults. Gracia~\cite{gracia2024allocating} embeds parity bits into invariant bits of BF16, making its effectiveness dependent upon data characteristics, thus limiting its applicability across different DNN models. SERA-Float~\cite{mishra2025sera} selectively protects floating-point fields using embedded parity, but it cannot guarantee reliable correction under certain single and double-bit fault scenarios.

To summarize, existing approaches either reduce computational overhead at the cost of limited protection capability or provide uniform ECC protection across all bits, which considerably increases computational overhead. To address these limitations, we propose UniCASE, which retains only the minimum set of computation-critical data bits required for accurate DNN execution. UniCASE uses a reduced number of parity bits to minimize approximation error and encoding/decoding cost while assigning protection strength according to bit criticality. As a result, UniCASE provides stronger fault protection with lower storage and computational overhead.

\section{Our Fault Model}
We assume faults occur in DNN parameters stored in SRAM/DRAM structures without ECC protection, reflecting commodity GPU settings where the lack of ECC support has been identified as a reliability concern \cite{haque2010hard}. We model soft errors as random bit flips and consider single-bit errors (SE), double-adjacent errors (DAE), and triple-adjacent errors (TAE), as prior studies show that single-bit upsets dominate memory faults, while technology scaling increases the likelihood of spatially local multi-cell upsets in adjacent memory cells \cite{hashimoto2019characterizing,bhuva2015multi}. In our experiments, SEs dominate the fault injection campaign, while DAE and TAE are injected with probabilities of 2\% and 1\%, respectively. 

\section{Floating-point block profiling}
\begin{figure}[t]
        \centering
\includegraphics[width=0.5\textwidth]{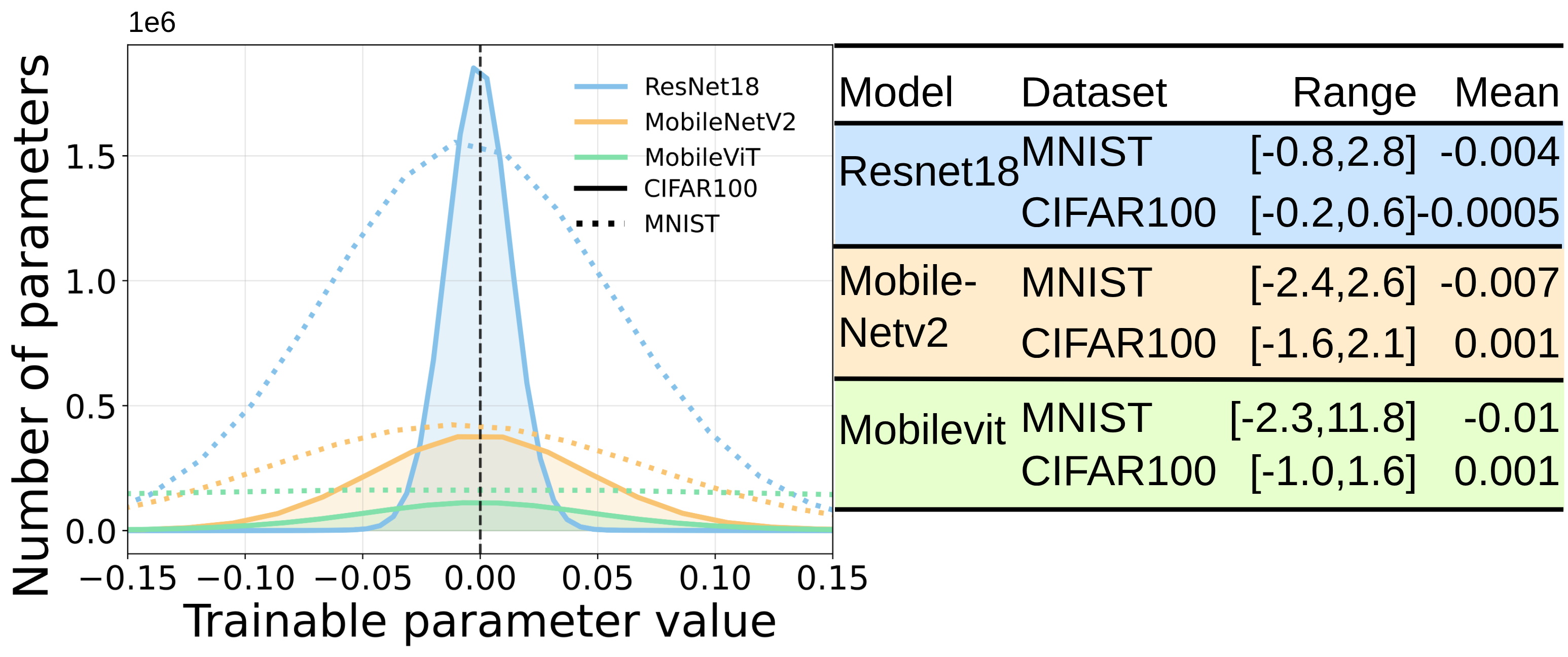}
    \caption{Distribution of trainable parameter values for ResNet18, MobileNetV2, and MobileViT trained on CIFAR-100 and MNIST}
    \label{fig:gaussian}
\end{figure}
\begin{figure}[t]
        \centering
\includegraphics[width=0.5\textwidth]{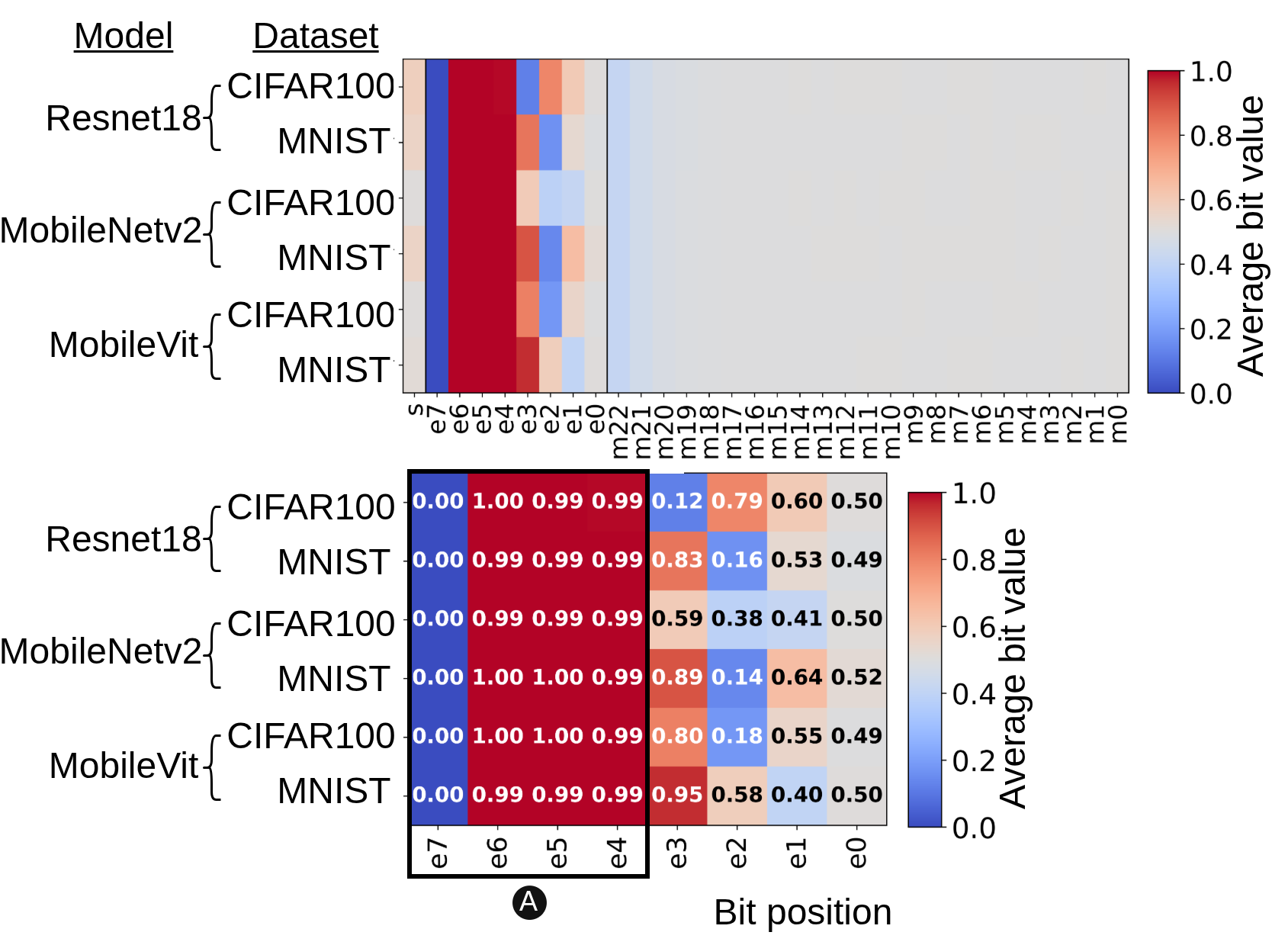}
    \caption{Average FP32 bit values of trainable parameters across ResNet18, MobileNetV2, and MobileViT trained on CIFAR-100 and MNIST}
    \label{fig:avg_bit}
\vspace{-3mm}
\end{figure}

\label{analysis}
To design a compact DNN data format, we first profile the numerical and bit-level behavior of trainable parameters. Across the evaluated DNNs, parameter values are concentrated around zero and approximately follow a Gaussian-like distribution, as shown in Figure \ref{fig:gaussian}, indicating that most parameters occupy a much smaller numerical range than the full IEEE floating-point space.

Floating-point formats consist of three fields: sign \(S\), exponent \(E\), and mantissa \(M\). For a floating-point format with \(n_E\) exponent bits and \(n_M\) mantissa bits, the exponent and mantissa fields are defined as
\(
E={e_{n_E-1}, e_{n_E-2}, \ldots, e_0},
\qquad
M={m_{n_M-1}, m_{n_M-2}, \ldots, m_0}.
\)
Within each field, the bit index represents significance. For instance, $e_{n_E-1}$ is the most significant exponent bit, $ e_0$ is the least significant exponent bit, $ m_{n_M-1} $ is the most significant mantissa bit, and $m_0$ is the least significant mantissa bit.

To quantify the stability of each floating-point field, we compute the average value of each bit position across all \(N\) parameters. For a bit \(b\in S\cup E\cup M\), the average bit value is defined as
\[
\bar{b}=\frac{1}{N}\sum_{i=1}^{N} b_i,
\]
where \(b_i\in\{0,1\}\) is the value of bit \(b\) in the \(i\)-th parameter. A bit with \(\bar{b}= 0\) or \(\bar{b}=1\) has no fluctuation in bit value, while \(\bar{b}\approx 0.5\) indicates high variability. Our profiling shows that the most significant exponent bits show very low fluctuations, as shown in the marked region \textcircled{A} in Figure \ref{fig:avg_bit}. \par
\begin{figure}[t]
        \centering
\includegraphics[width=0.5\textwidth]{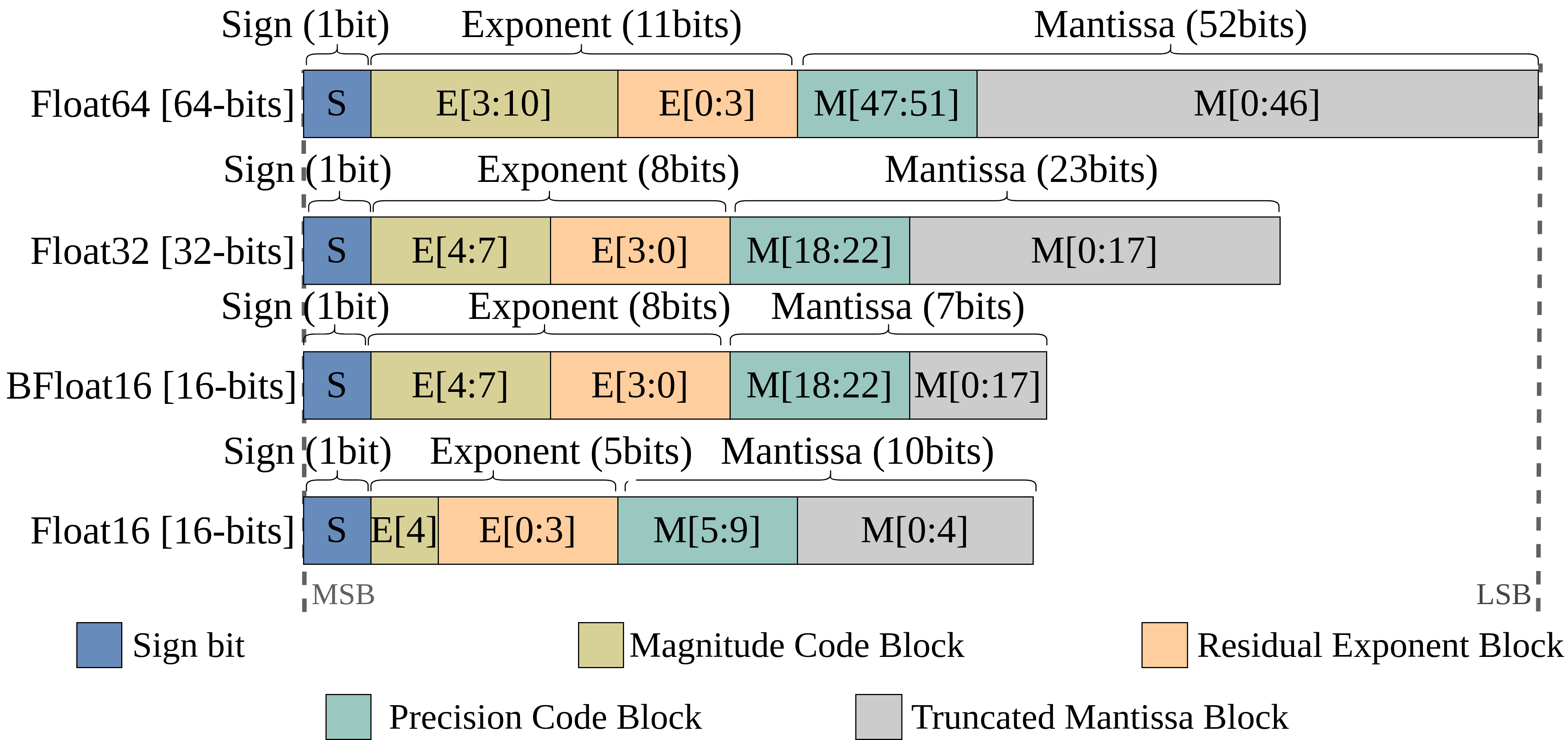}
    \caption{Classification of FP64, FP32, FP16, and BFloat16 into the sign bit, Magnitude Code Block, Residual Exponent Block, Precision Control Block, and Truncated Mantissa Block}
    \label{fig:block_class}
\end{figure}

 We call these most significant exponent bits the \textit{Magnitude Code Block} $B_{MC}$. The remaining exponent bits form the \textit{Residual Exponent Block} $B_{RE}$. For the mantissa, we retain only the five most significant bits,
called the \textit{Precision Control Block} $B_{PC}$, since lower mantissa bits have reduced numerical impact and DNNs are generally tolerant to small rounding errors \cite{gupta2015deep}. To set the length of each block and to verify that the proposed block definitions are consistent across floating-point formats, we compute a cross-format bit similarity score. For each model, trainable weights are converted to FP64, FP32, FP16, and BFloat16 and decomposed into sign, exponent, and mantissa fields. For a selected bit position \(b\), the agreement score is
\[
A_b=\frac{1}{N}\sum_{i=1}^{N}
\mathbb{I}\left(b_i^{\mathrm{FP64}}=b_i^{\mathrm{FP32}}=b_i^{\mathrm{FP16}}=b_i^{\mathrm{BF16}}\right),
\]
where \(N\) is the number of parameters. We select the sign bit, five exponent bits, and eight mantissa bits based on the smallest field lengths among the evaluated formats. The most significant exponent bits are excluded from this comparison since their interpretation depends on format-specific exponent length and bias.
\begin{figure}[t]
        \centering
\includegraphics[width=0.5\textwidth]{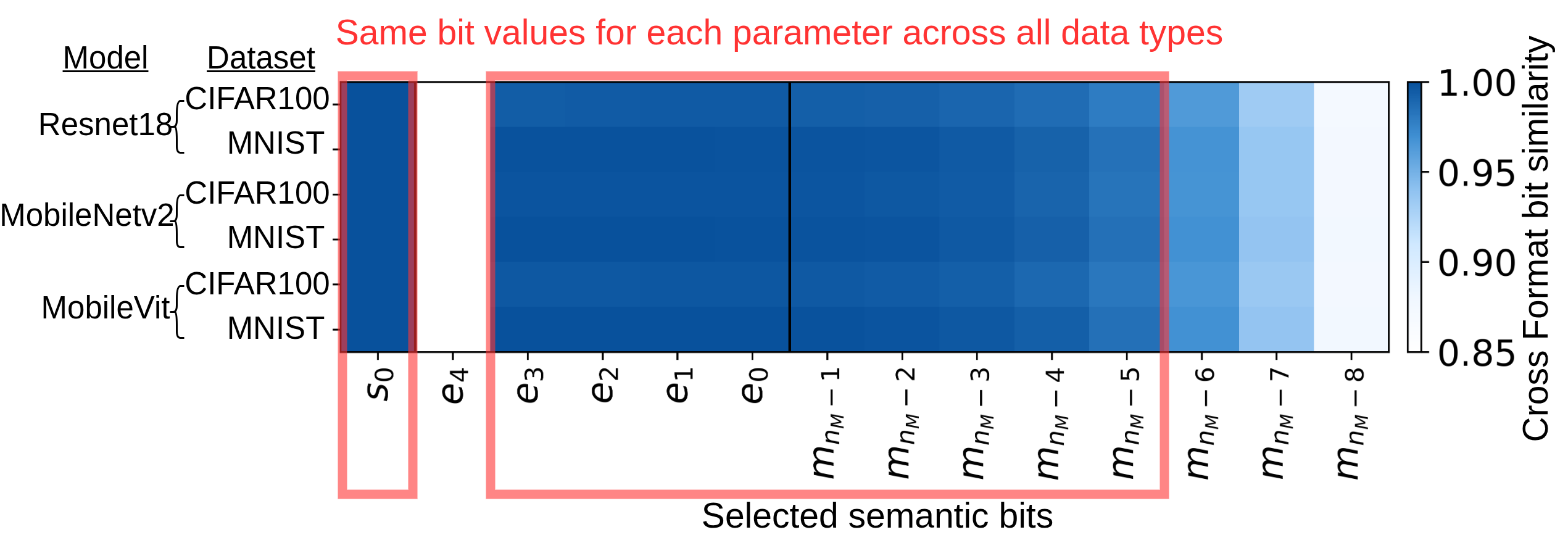}
    \caption{Cross-format bit similarity among FP64, FP32, FP16, and BFloat16}
    \label{fig:cross_data}
    \vspace{-3mm}
\end{figure} 
Figure \ref{fig:cross_data} shows that the sign bit, the four least significant exponent bits, and the five most significant mantissa bits of a given parameter remain highly consistent across FP32, FP64, FP16, and BFloat16. We therefore use these stable bit regions as data bits in UniCASE. This cross-format similarity enables parameters stored in different floating-point formats to be directly mapped into our unified data type. Figure \ref{fig:block_class} shows the block classification for FP64, FP32, FP16 and BFloat16.

\section{Our Novel UniCASE}
\begin{figure}[t]
        \centering
\includegraphics[width=0.5\textwidth]{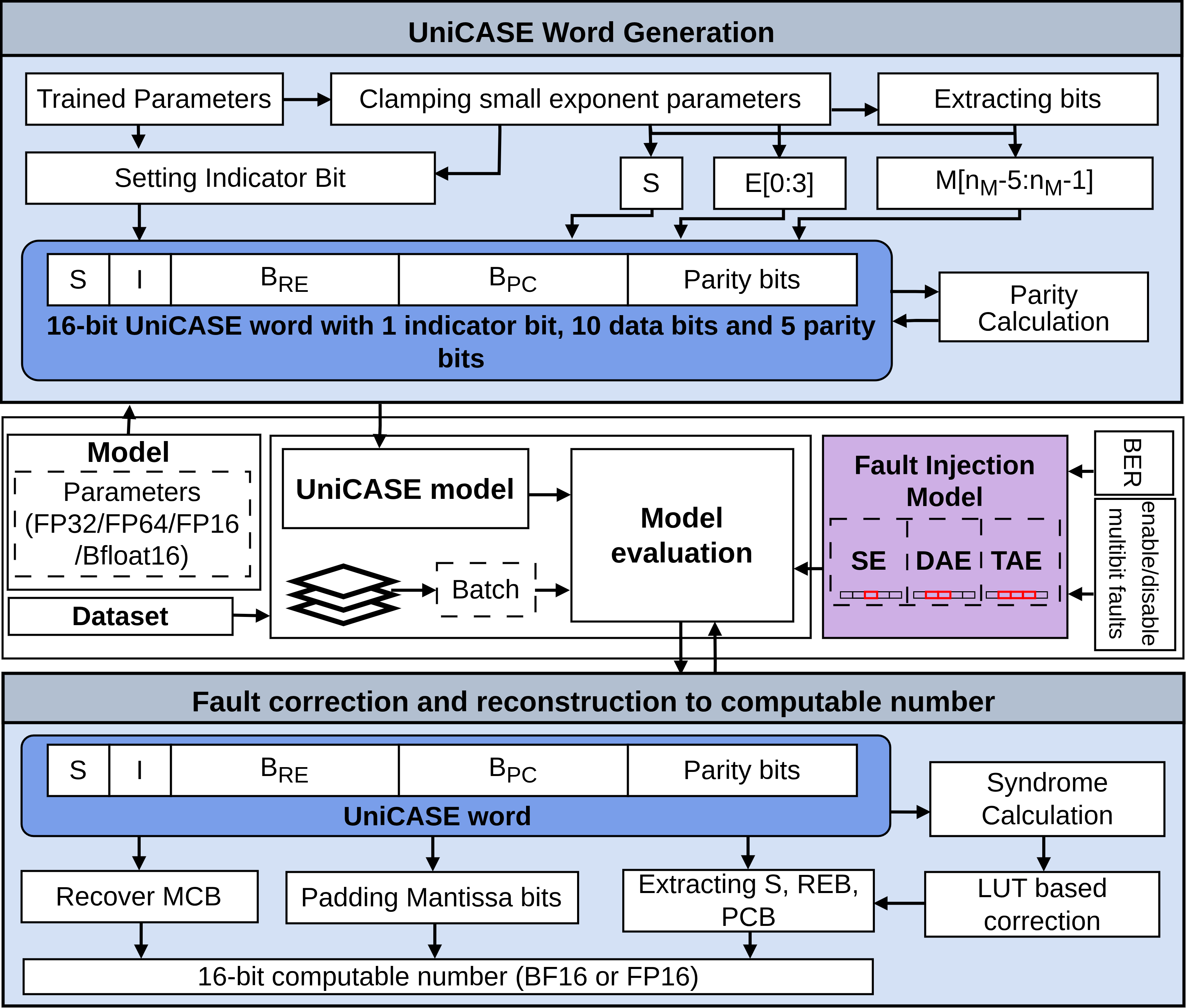}
    \caption{System overview of the proposed UniCASE}
    \label{fig:system}
    \vspace{-4mm}
\end{figure}

Figure~\ref{fig:system} shows the system overview of UniCASE. DNN's pretrained parameters are first converted into the UniCASE representation, which retains the essential data bits identified in Section~\ref{analysis} and embeds parity bits for error protection. During inference, UniCASE applies selective SEC/DAEC/TAEC correction to the protected word and reconstructs the corrected value into an executable numerical format for accurate and reliable DNN computation.
\subsection{UniCASE Word Generation}
The fluctuation in \textit{Magnitude Code Block} is caused by very small exponent values or large exponent values (values at tail of the Gaussian curve). Therefore, parameters with very small magnitudes, which have a negligible impact on DNN execution, are clamped to zero to reduce bit-level fluctuation:
\[
\theta_i =
\begin{cases}
0, & |\theta_i| < 10^{-4},\\
\theta_i, & \text{otherwise}.
\end{cases}
\]
where \(\theta_i\) is the \(i\)-th trainable parameter of a DNN model. For larger exponent values, floating-point parameters are divided into two ranges, \(2^{-15} \le |\theta_i| < 2^{1}\) and \(2^{1} \le |\theta_i| < 2^{17}\), because each range corresponds to a distinct \(B_{MC}\). To store this information, we add an indicator bit \(I\),
\[
I =
\begin{cases}
0, & 2^{-15} \le |\theta_i| < 2^{1} \\
1, & 2^{1} \le |\theta_i| < 2^{17}
\end{cases}
\] In UniCASE, \(B_{MC}\) is replaced by indicator bit \(I\), which distinguishes between the two exponent regions. The data bits of compressed representation consist of sign bit $S$, indicator bit $I$,  \(B_{RE}\) and \(B_{PC}\).

\subsection{Criticality-Aware Multi-Level ECC}
\label{CriticalityAware}
\subsubsection{Bit-Level Criticality Classification}
\label{low_parity}
\begin{figure}[t]
        \centering
\includegraphics[width=0.5\textwidth]{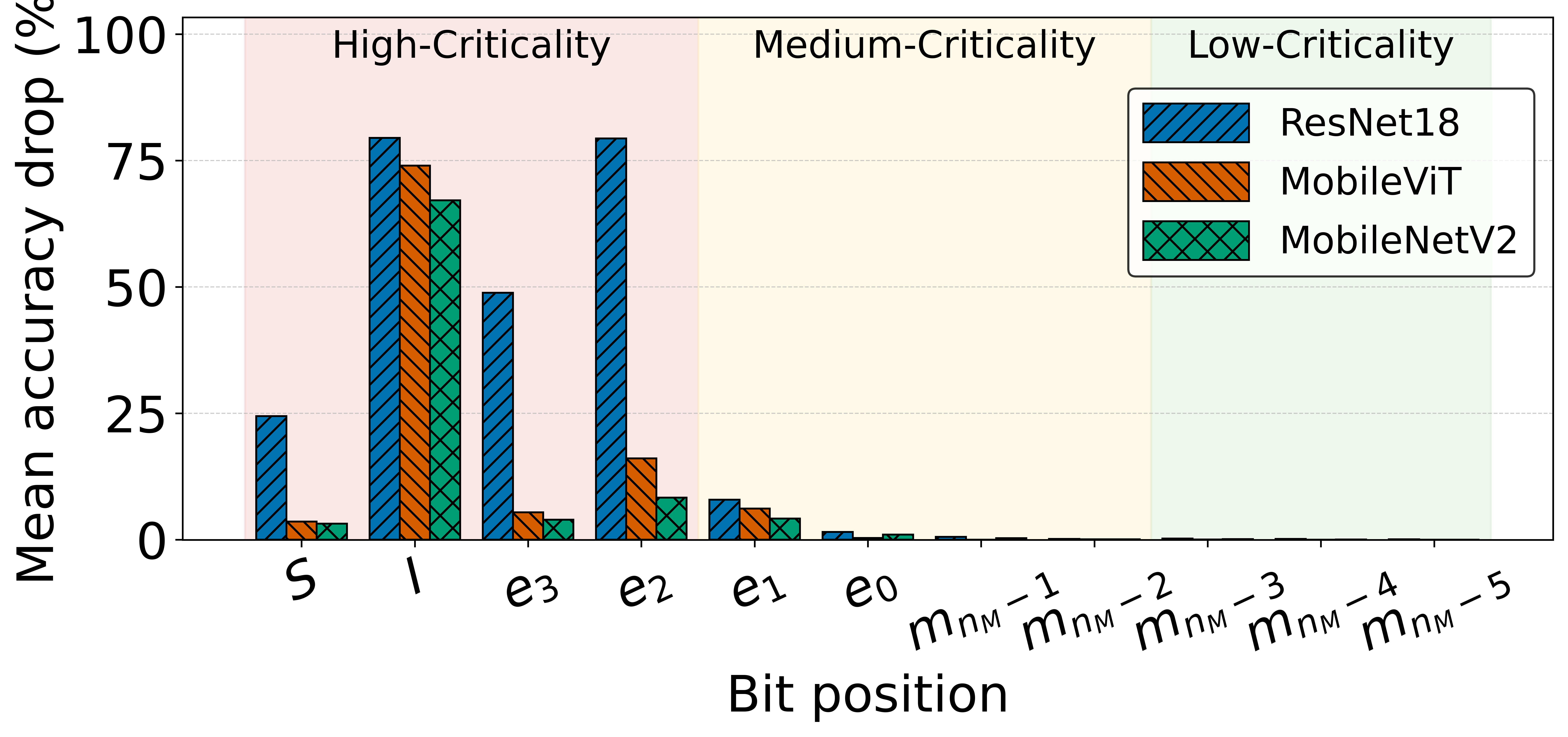}
    \caption{Bit-level criticality classification of UniCASE. Bits are classified into high, medium, and low criticality levels based on their impact on final model accuracy under fault injection.}
    \label{fig:criticality}
\end{figure}

The retained UniCASE data bits are first classified into high-, medium-, and low-criticality groups using bit-level fault injection. Faults are injected at each bit position, and the resulting accuracy degradation is measured, as shown in Figure~\ref{fig:criticality}. Bits \(S\), \(I\), \(e_3\), and \(e_2\) are classified as high-criticality because faults in these positions can cause accuracy loss of up to \(80\%\). Bits \(e_1\), \(e_0\), \(m_{22}\), and \(m_{21}\) are classified as medium-criticality due to moderate accuracy degradation, while \(m_{20}\), \(m_{19}\), and \(m_{18}\) are classified as low-criticality because faults at these positions have negligible impact on model accuracy, as shown in Figure~\ref{fig:criticality}. Parity bits are also treated as high-criticality bits because parity faults can generate incorrect syndromes and trigger false correction events.
\subsubsection{Designing Multi-level ECC}
UniCASE has a fixed word size of \(n=16\) bits, including both data and parity bits. To support SEC/DAEC/TAEC within a fixed 16-bit word, we define the syndrome budget according to the number of available parity bits. As shown in Tables~\ref{tab:parity_coverage} and~\ref{tab:syndrome_requirement_unicase}, full SEC-DAEC-TAEC protection over all 16 bit positions requires separate syndromes for the no-fault case, all single-bit errors, all double-adjacent errors, and all triple-adjacent errors. Therefore, the required number of syndromes is:
\[
N_{\text{full}}
= 1 + N_{\text{SEC}} + N_{\text{DAEC}} + N_{\text{TAEC}}=3n-2=46
\] Four parity bits provide only 16 syndromes, which is insufficient for full 16-bit SEC and leaves no syndrome budget for DAEC or TAEC. UniCASE therefore uses \(r=5\) parity bits, providing \(
N_{\text{syn}} = 2^r = 2^5
\) syndromes, and allocates them to the most accuracy-critical SE, DAEC, and TAEC patterns.
Since
\(
2^5<46,
\)
five parity bits are insufficient to provide uniform full SEC-DAEC-TAEC protection over the complete 16-bit word.

To satisfy the limited syndrome budget, we use criticality-aware selective protection. Let the high-criticality set be
$\mathcal{H}=
\{S,I,e_3,e_2,p_0,\\p_1,p_2,p_3,p_4\},
$
where \(S\), \(I\), \(e_3\), and \(e_2\) are high-criticality data bits and \(p_0,\ldots,p_4\) are parity bits. We reserve 28 syndromes for SEC-DAEC-TAEC recovery of the high-criticality set. 
The four remaining syndromes are assigned to single-bit correction of the four medium-criticality bits. The low-criticality bits are left unprotected.

\begin{table}[t]
\centering
\caption{Protection coverage for a 16-bit word under SE-DAEC-TAEC correction.}
\label{tab:parity_coverage}
\resizebox{\columnwidth}{!}{
\begin{tabular}{c c c c}
\toprule
\textbf{Parity Bits} & \textbf{Syndromes} & \textbf{Fault Patterns Covered} & \textbf{Coverage} \\

\midrule
4 & $2^4=16$ & $16/46$ & $34.7\%$ \\
5 & $2^5=32$ & $32/46$ & $69.5\%$ \\
6 & $2^6=64$ & $46/46$ & $100\%$ \\
\bottomrule
\end{tabular}
}
\end{table}
\begin{table}[t]
\centering
\caption{Syndrome requirement for full 16-bit SEC-DAEC-TAEC protection and the selective protection used in UniCASE.}
\label{tab:syndrome_requirement_unicase}
\resizebox{\columnwidth}{!}{
\begin{tabular}{lccc}
\toprule
\textbf{Fault} & \textbf{Full 16-bit} & \textbf{UniCASE} & \textbf{UniCASE} \\

\textbf{Model} & \textbf{Requirement} & \textbf{Coverage} & \textbf{Protection} \\
\midrule
No-fault & 1 syndrome & 1 syndrome & x \\
\midrule
SE   & 16 syndromes & 13 protected bits & High-criticality \\
 &  & & + medium-criticality\\
 &  & & + parity bits \\
 \midrule
DAEC & 15 syndromes & 9 protected bits & High-criticality \\
 &  & & + parity bits \\
 \midrule
TAEC & 14 syndromes & 9 protected bits & High-criticality \\
 &  & & + parity bits \\
\midrule
\textbf{Total} & \textbf{46 syndromes} & \textbf{32 syndromes} & \textbf{5-bit selective ECC} \\
\bottomrule
\end{tabular}
}
\end{table}
\begin{figure}[t]
        \centering
\includegraphics[width=0.5\textwidth]{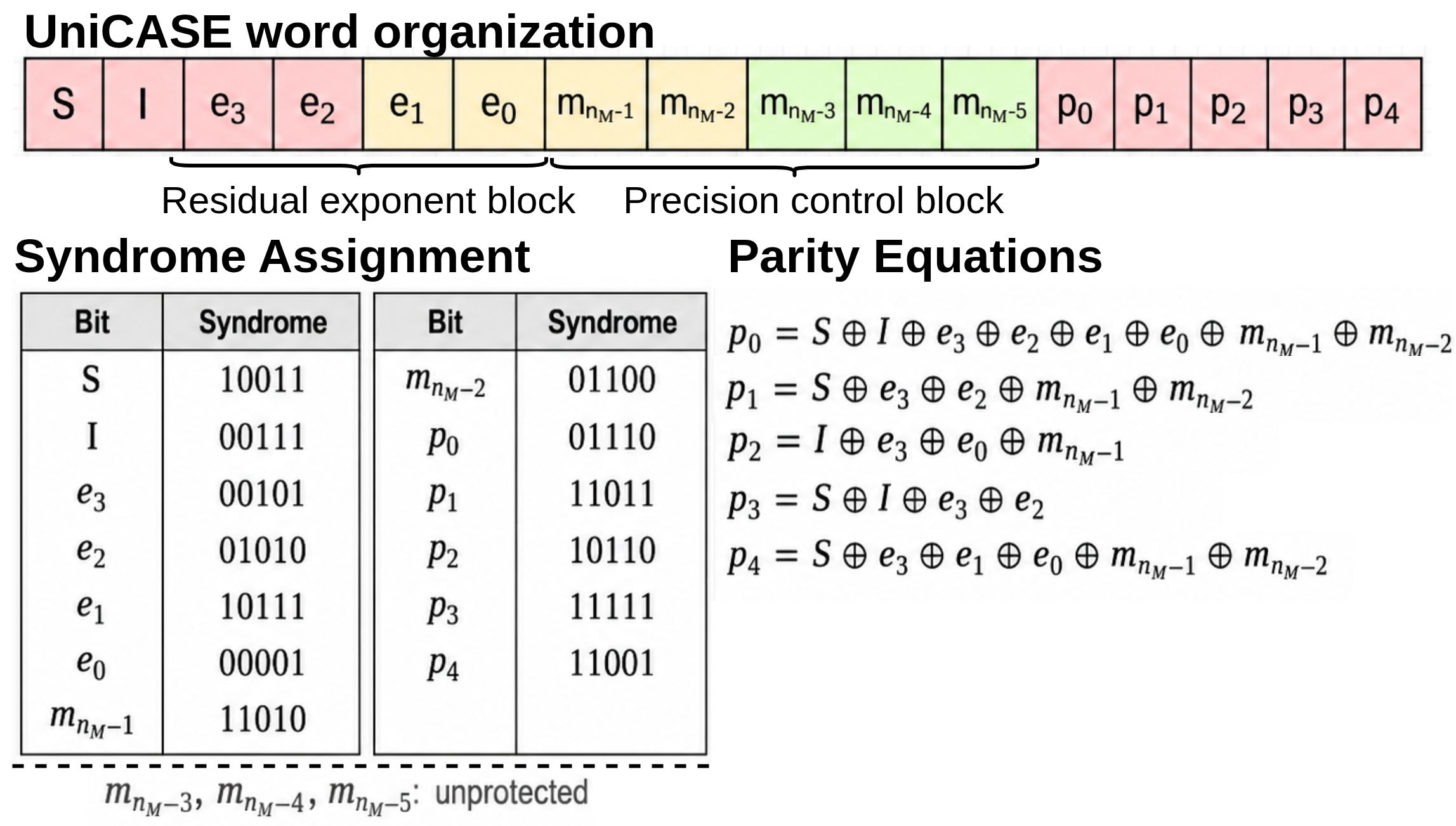}
    \caption{UniCASE word organization with the assigned syndromes and parity equations used for selective bit-level error protection.}
    \label{fig:syndrome}

\end{figure}
\begin{table}[t]
\centering
\caption{Accuracy comparison without fault injection for FP32, SERA-Float~\cite{mishra2025sera}, Gracia~\cite{gracia2024allocating}, and UniCASE.}
\label{tab:accuracy_comparison}
\renewcommand{\arraystretch}{1.}
\large
\setlength{\tabcolsep}{2.5pt}
\resizebox{\columnwidth}{!}{%

\begin{tabular}{lcccccc}
\toprule
 & \multicolumn{3}{c}{\textbf{CIFAR100}} 
 & \multicolumn{3}{c}{\textbf{MNIST}} \\
\cmidrule(lr){2-4} \cmidrule(lr){5-7}
\textbf{Component} 
& \textbf{ResNet18} & \textbf{MobileNetv2} & \textbf{Mobilevit} 
& \textbf{ResNet18} & \textbf{MobileNetv2} & \textbf{Mobilevit}  \\

\midrule

FP32    & \textbf{80.3} & \textbf{79.6} & \textbf{76.6} & \textbf{99.2} & \textbf{98.8} & \textbf{99.3} \\

Gracia\cite{gracia2024allocating}(16,10)+3& 78.5 & 1.0 & 48.0 & 99.2 & 11.3& 11.3 \\
Gracia\cite{gracia2024allocating}(16,10)+4& 1.1 & 0.9 & 1.0 & 5.8 & 24.4 & 11.35 \\
Gracia\cite{gracia2024allocating}(16,10)+5& 0.9 & 1.0 & 1.0 & 10.3 & 10.1 & 0.0 \\
SERA-Float\cite{mishra2025sera}& \textbf{80.3} & 79.5 & \textbf{76.6} & \textbf{99.2} & \textbf{98.8} & \textbf{99.3} \\
UniCASE (Ours)& \textbf{80.3} & 79.2 & 76.5 & \textbf{99.2} & \textbf{98.8} & \textbf{99.3} \\
\bottomrule
\end{tabular}%
}
\footnotesize{All values are reported as accuracy (\%).}
\end{table}

\subsubsection{Syndrome Assignment Search}
\begin{figure*}[t]
        \centering
\includegraphics[width=1.0\textwidth]{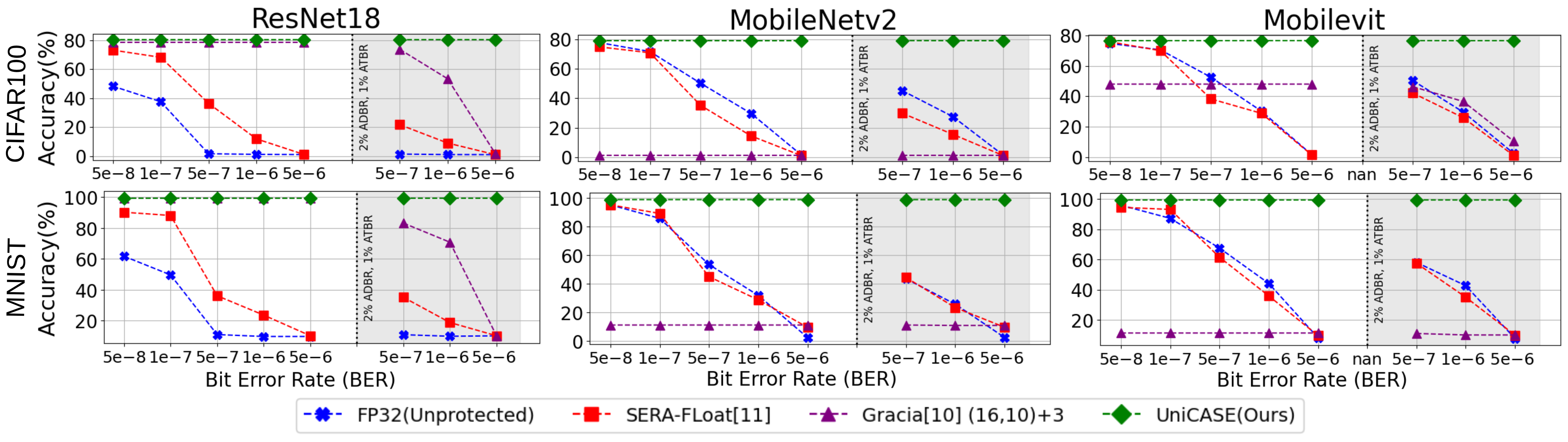}
    \caption{Accuracy comparison of unprotected FP32, SERA-Float~\cite{mishra2025sera}, Gracia~\cite{gracia2024allocating}, and UniCASE on ResNet18, MobileNetV2, and MobileViT trained on CIFAR-100 and MNIST datasets under different bit-error rates (BER).}
    \label{fig:results}
\end{figure*}
After defining the criticality-aware protection targets, we construct the 5-bit syndrome space and derive the parity equations for the ECC encoder and decoder. Since five parity bits provide only \(2^5\) possible syndromes, UniCASE cannot assign unique syndromes to all full SEC/DAEC/TAEC error patterns. Let \(H(b)\) denote the nonzero syndrome assigned to a protected bit \(b\). For an error pattern \(\mathcal{E}\), the syndrome is computed as
\[
S(\mathcal{E})=\bigoplus_{b\in \mathcal{E}\cap \mathcal{P}} H(b),
\]
where \(\mathcal{P}\) is the set of protected bits. The corresponding correction action is defined only over protected bits.

We formulate syndrome assignment as a bounded randomized min-conflict search~\cite{minton1992minimizing}. The search starts with unique nonzero syndromes for all protected bits and evaluates all required SE patterns, selected DAEC patterns, and selected TAEC patterns. A collision is harmful only if two patterns produce the same syndrome but require different correction actions and aliases with the same correction action are allowed. The objective is
\(
\min_H C(H),
\)
where \(C(H)\) is the number of harmful collisions. At each iteration, one protected-bit syndrome is mutated while preserving uniqueness and nonzero values. The assignment is accepted if it reduces \(C(H)\), or probabilistically accepted with a decreasing temperature to escape local minima. Multiple random restarts are used until a collision-free assignment is found. The final assignment is additionally checked to prevent medium-criticality DAEC/TAEC patterns from aliasing with high-criticality data-bit corrections, while same-action parity-bit aliases are allowed.

After a valid assignment is found, the parity equations are derived over \(GF(2)\) as
\[
\mathbf{p}=P^{-1}A\mathbf{d},
\]
where \(P\) is the parity submatrix, \(A\) is the data-bit submatrix, and \(\mathbf{d}\) and \(\mathbf{p}\) are the data and parity vectors, respectively. The resulting decoder table maps each valid syndrome to its correction action. The final syndrome assignment and parity equations used in UniCASE are shown in Figure~\ref{fig:syndrome}.
\subsection{Computable value reconstruction}
After correction, the word is reconstructed into a standard BFloat16 or FP16 value. The indicator bit recovers the \textit{Magnitude Code Block}, while the discarded least significant mantissa bits are added by fixed padding.
\section{Evaluation}
\subsection{Fault-injection Model}
For fault-injection evaluation, we target all model parameters stored in memory. For the original FP32 model, we evaluate BER values ranging from \(10^{-8}\) to \(10^{-6}\). For UniCASE, the same absolute number of faults is injected for each BER point, without scaling by the smaller word size, to ensure a fair comparison across methods. For each test batch, the model is reset, and a new set of random bit flips is injected, creating a dynamic fault pattern throughout evaluation. The BER is defined as
\begin{equation}
\text{BER}=\frac{n_f}{N},
\label{eq:ber}
\end{equation}
where \(n_f\) is the number of injected faults and \(N\) is the total number of model bits.

We conduct two sets of fault-injection experiments. In the first set of experiments, only single-bit errors (SEs) are injected to evaluate the baseline resilience of each format against isolated soft errors. In the second set, we inject single-bit errors (SE), double-adjacent errors (DAE), and triple-adjacent errors (TAE), where DAE events account for (2\%) of the total fault count, and the number of TAE events is half of the DAE count. Since adjacent multi-bit faults are rare, we enforce at least one DAE and one TAE per campaign to ensure that robustness against these cases is evaluated. This fault model captures both isolated soft errors and spatially correlated adjacent multi-bit upsets.

\subsection{Model and Datasets}To evaluate UniCASE on image classification, we conduct experiments on convolutional neural networks (CNNs) and vision transformers using CIFAR100 and MNIST datasets. For CNN architectures, we select ResNet18 \cite{he2016deep} and MobileNetv2 \cite{sandler2018mobilenetv2}, and for vision-transformer, we use Mobilevit \cite{mehta2021mobilevit}. All models are initialized with pretrained weights, and the classification layer is replaced to match the number of classes. Moreover, each model is fine-tuned for 50 epochs on both datasets using cross-entropy loss as the loss function with the Adam optimizer and a batch size of 64. We use PyTorch and the timm library to load and fine-tune the models.
\section{Results and Discussion}
\subsection{Accuracy loss due to approximation}
\label{acc_loss}

Since UniCASE approximates exponent values and drops selected mantissa bits to incorporate parity bits, we first evaluate the accuracy of DNNs using UniCASE in the absence of faults. We compare our format against FP32, used as the golden baseline, as well as Gracia \cite{gracia2024allocating} (16,10)+3, Gracia \cite{gracia2024allocating} (16,10)+4, Gracia \cite{gracia2024allocating} (16,10)+5, and SERA-Float \cite{mishra2025sera}. The results are reported in Table~\ref{tab:accuracy_comparison}. As shown, UniCASE introduces zero to less than \(1\%\) accuracy degradation due to approximation error, and it achieves accuracy comparable to FP32 and SERA-Float while using only half the memory, as shown in Figure \ref{fig:size_comp}. \par In contrast, Gracia \cite{gracia2024allocating} shows limited generalizability and suffers from a significant accuracy drop in the Gracia \cite{gracia2024allocating} (16,10)+4 and Gracia \cite{gracia2024allocating} (16,10)+5 configurations. This degradation occurs because Gracia assumes that the most significant bits remain invariant, which does not always hold, as illustrated in Figure~\ref{fig:avg_bit}. However, Gracia~\cite{gracia2024allocating} \((16,10)+3\) performs comparatively well on ResNet-18 due to the model's parameter distribution.

\subsection{Robustness Under Fault Injection}

Figure~\ref{fig:results} compares the fault resilience of UniCASE with SERA-Float~\cite{mishra2025sera} and Gracia~\cite{gracia2024allocating} \((16,10)+3\). We use \((16,10)+3\) as the representative Gracia \cite{gracia2024allocating} configuration because it is the only evaluated variant that remains functional on ResNet18, while the other variants show limited generalizability across DNN models. The results show that UniCASE consistently outperforms both SERA-Float \cite{mishra2025sera} and Gracia \cite{gracia2024allocating} under injected faults.\par SERA-Float suffers significant accuracy degradation even under single-bit faults, despite using \(2.8\times\) the number of parity and control bits compared with UniCASE. Gracia \cite{gracia2024allocating} \((16,10)+3\) provides partial protection but does not support adjacent triple-bit error correction, its accuracy degrades as adjacent triple-bit faults increases. Moreover, Gracia \cite{gracia2024allocating} requires more parity bits than UniCASE while providing weaker adjacent multi-bit protection.\par In contrast, UniCASE maintains substantially higher classification accuracy under moderate to high fault rates, including adjacent double and triple-bit fault scenarios. This demonstrates that criticality-aware selective ECC can provide stronger model-level resilience with fewer parity bits.

\subsection{Memory and Computational Overhead}
\label{comparison}
\begin{figure}[t]
        \centering
\includegraphics[width=0.5\textwidth]{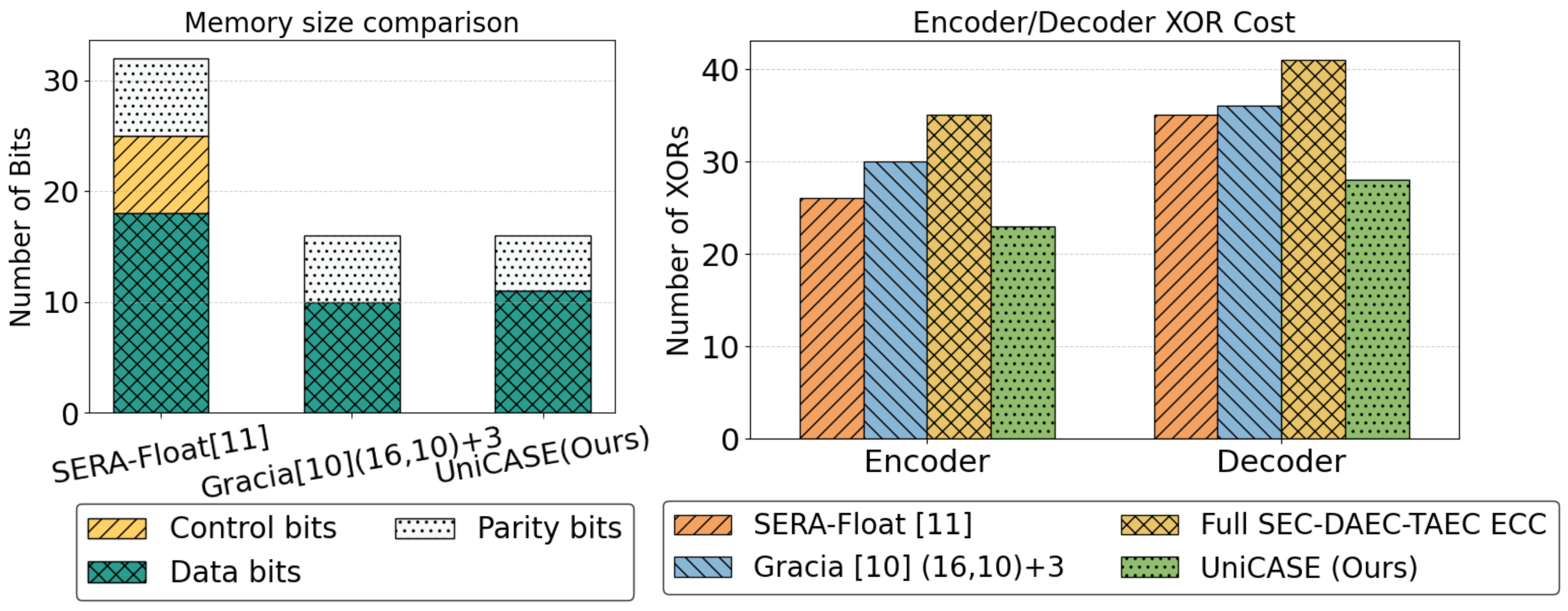}
    \caption{Memory and XOR cost comparison of SERA-Float~\cite{mishra2025sera}, Gracia~\cite{gracia2024allocating}, and UniCASE}
    \label{fig:size_comp}
\end{figure}

Figure~\ref{fig:size_comp} shows that both UniCASE and Gracia~\cite{gracia2024allocating} use only half the storage bits required by SERA-Float \cite{mishra2025sera}. However, compared with Gracia~\cite{gracia2024allocating}, UniCASE uses fewer parity bits to reduce both approximation loss and computational overhead. As discussed in Section \ref{low_parity}, allocating more parity bits within a fixed-width format reduces the number of available data bits, which can increase approximation error. In addition, each extra parity bit introduces additional XOR operations during encoding and decoding. This overhead becomes significant when millions of DNN parameters must be encoded, decoded, and corrected during inference. Moreover, for a similar protection level of full SEC-DAEC-TAEC, UniCASE reduces XOR cost by 30\%, as shown in Figure~\ref{fig:size_comp}.
\subsection{Generalizability of UniCASE}
\begin{table}[t]
\centering
\caption{Generalizability of UniCASE across NLP models and LLMs.}
\label{tab:unicase_generalization}
\resizebox{\columnwidth}{!}{
\begin{tabular}{lcccc}
\toprule
\multirow{2}{*}{\textbf{Model}} 
& \multicolumn{2}{c}{\textbf{GLUE-SST2 \cite{wang2018glue}}} 
& \multicolumn{2}{c}{\textbf{Yelp-Polarity \cite{zhang2015character}}} \\
\cmidrule(lr){2-3} \cmidrule(lr){4-5}
& \textbf{FP32} & \textbf{UniCASE} 
& \textbf{FP32} & \textbf{UniCASE} \\
\midrule
DeBERTa-v3-small \cite{he2021debertav3} & 100 & 100& 96& 96 \\
Qwen2.5 \cite{hui2024qwen2}          & 83  & 83 & 86& 86 \\
RoBERTa \cite{liu2019roberta}         & 93 & 93 & 83& 83 \\
SmolLM2 \cite{allal2025smollm2}         & 43  & 43 & 46& 46 \\
TinyLlama \cite{zhang2024tinyllama}       & 43  & 43  & 46 & 46\\
\bottomrule
\end{tabular}
}
\footnotesize{All values are reported as accuracy (\%).}
\vspace{-3mm}
\end{table}
UniCASE supports the range \(2^{-15} \le |\theta| < 2^{17}\), motivated by the observation that trained DNN parameters follow a roughly Gaussian-like distribution with only a small fraction of values in the tails. This behavior is consistent with common DNN training configurations \cite{he2015delving,ioffe2015batch}. \par To evaluate the generalizability of UniCASE, we further apply UniCASE to a diverse set of NLP models and large language models (LLMs) without performing model parameter profiling. Each model is directly converted from FP32 to UniCASE, and the resulting accuracy is compared with the original FP32 baseline. As shown in Table~\ref{tab:unicase_generalization}, UniCASE incurs zero to negligible accuracy degradation across all evaluated models. These results indicate that UniCASE is not limited to a specific model architecture, dataset, or task. Instead, UniCASE provides a numerical range that is broadly representative of trained neural-network parameters, enabling conversion with negligible accuracy degradation across both conventional DNNs and LLMs.

\section{Conclusion}
In this paper, we present UniCASE, a unified compressed data format with criticality-aware selective ECC for reliable DNN parameter protection. UniCASE protects DNN parameters against single-bit faults, double-adjacent faults, and triple-adjacent faults by assigning different protection strengths to different bit groups according to their impact on model accuracy. By retaining only the data bits essential for accurate DNN execution and using fewer parity bits than conventional uniform protection, UniCASE reduces approximation error, storage redundancy, and encoding/decoding cost. Experimental results demonstrate that UniCASE offers stronger soft-error resilience than state-of-the-art protection methods while minimizing computational cost. Furthermore, UniCASE does not require model-specific profiling or retraining and generalizes across vision models, NLP models, and LLMs.

\bibliographystyle{IEEEtran}
\bibliography{references}

\end{document}